\documentclass[aps,preprint]{revtex4}
\usepackage{epsfig}
\usepackage{amsmath}
\usepackage{epsfig}
\usepackage{float}
\usepackage{dcolumn} 
\usepackage{braket}

\begin{document}
\title
{Coalescing versus merging of energy levels in one-dimensional potentials}    
\author{Zafar Ahmed$^1$, Sachin Kumar$^2$, Achint Kumar$^3$, Mohammad Irfan$^4$} 
\affiliation{$~^1$Nuclear Physics Division, Bhabha Atomic Research Centre, Trombay, Mumbai-85, India\\
$~^2$Theoretical Physics Section, Bhabha Atomic Research Centre, Trombay, Mumbai-85, India\\	
$~^3$Department of Physics, Birla Institute of Technology \& Science, Pilani, Goa, 403726, India\\
$~^4$Department of Physics, Indian Institute of Science Education and Research, Bhopal,  462066, India}
\email{1:zahmed@barc.gov.in,2:sachinv@barc.gov.in, 3:achint1994@gmail.com, 4:mohai@iiserb.ac.in}
\date{\today}
\begin{abstract}
\noindent
The  sub-barrier pairs of energy levels of a Hermitian  one-dimensional symmetric double-well potential are known to merge into one, if the inter-well distance ($a$) is increased slowly. The energy at which the doublets merge are the ground state eigenvalues of independent 
 ($\epsilon_0$). We show that if the double-well is perturbed  mildly by a complex PT-symmetric potential, the merging of levels turns into the coalescing of two levels at an exceptional point $a=a_*$. For $a>a_*$, the real part of complex-conjugate eigenvalues coincides with  $\epsilon_0$ again. This is an interesting and rare connection between the two phenomena in two domains: Hermiticity and complex PT-symmetry. 
\end{abstract}
\maketitle
In one dimensional quantum mechanics there is  one to one correspondence between eigenvalues and eigenstates, there is an absence of degeneracy. So when a parameter of the Hamiltonian is varied slowly curves can not cross
but they can come quite close and then diverge from each other (Avoided Crossing). Crossings and avoided crossings of levels is commonly observed in  the spectra of two or three dimensional systems.  Mostly, in one dimensional systems if a parameter of the potential is varied slowly, the eigenvalues increase or decrease monotonically. For particle in  an infinitely deep well of width $a$, $E_n(= \frac{n^2 \pi^2 \hbar^2}{2ma^2}$) decrease as function of $a$. For harmonic oscillator potential, $E_n=(n+1/2)\hbar\omega$ increase linearly as function of the frequency parameter $\omega$. 

Two levels coming very close may either display merging of two levels or their avoided crossing. The former is well known to occur in symmetric double-well potentials wherein the sub-barrier doublets of energy levels merge [1,2] into the levels of the independent wells when the inter-well distance is increased slowly. On the other hand AC is observed rather rarely in one dimensional systems. Recently, it has been shown [2,3] that in double-well potentials if the width or depth of the potential is varied slowly very interesting level crossings can be observed. Notably the double-well becomes asymmetric.

For one-dimensional non-Hermitian Hamiltonions, it is known that two complex  eigenvalues may become real at one special value of the parameter($\lambda= \lambda_*))$ of the potential after this point these two eigenvalues may again be complex. Such special values of the parameter are called Exceptional Point (EP) [4]. More interestingly, when a potential PT-symmetric (invariant under the joint action of Parity: $x\rightarrow -x$ and Time-reversal ($i \rightarrow -i$), the two discrete eigenvalues make a transition from real  to complex-conjugate or {\it vice versa}. For instance, in the complex PT-symmetric potential: $V(x)=-V_1 \mbox{sech}^2x +i|V_2| \mbox{sech}x \tanh x$, $|V_2|=V_1+1/4=V_c ~(2m=1=\hbar^2)$ is the EP of  this potential when $|V_2| \le V_c$ eigenvalues are real, otherwise these are complex conjugate pairs[5].
\begin{figure}[t]
\centering
\includegraphics[width=4 cm,height=5 cm]{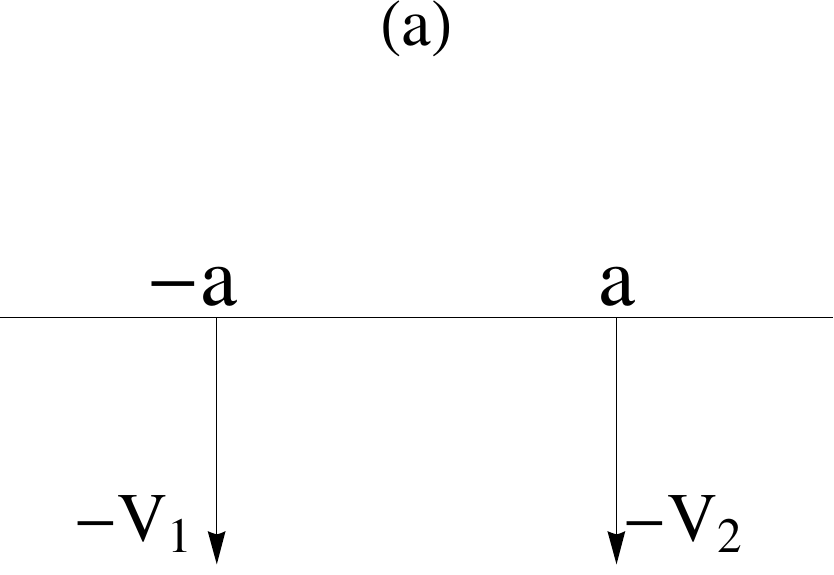}
\hskip .5cm
\includegraphics[width=4 cm,height=5 cm]{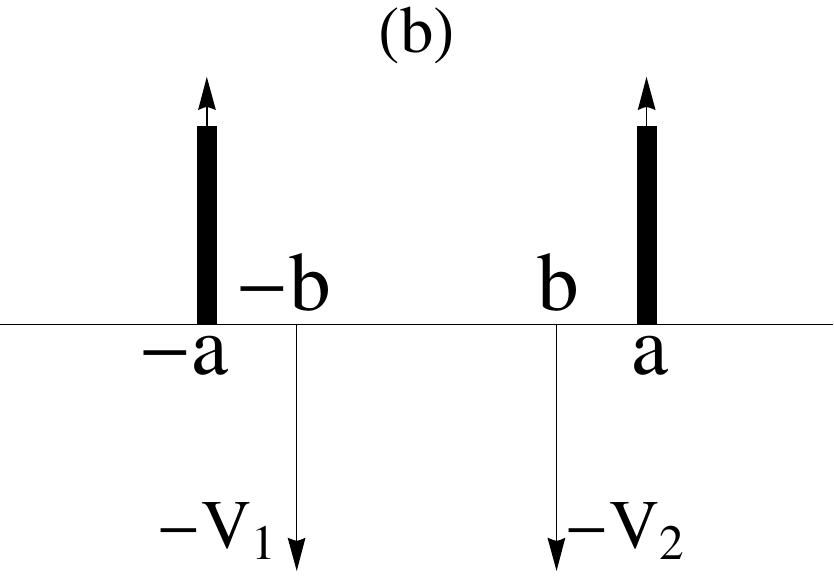}
\hskip .5cm
\includegraphics[width=4 cm,height=5 cm]{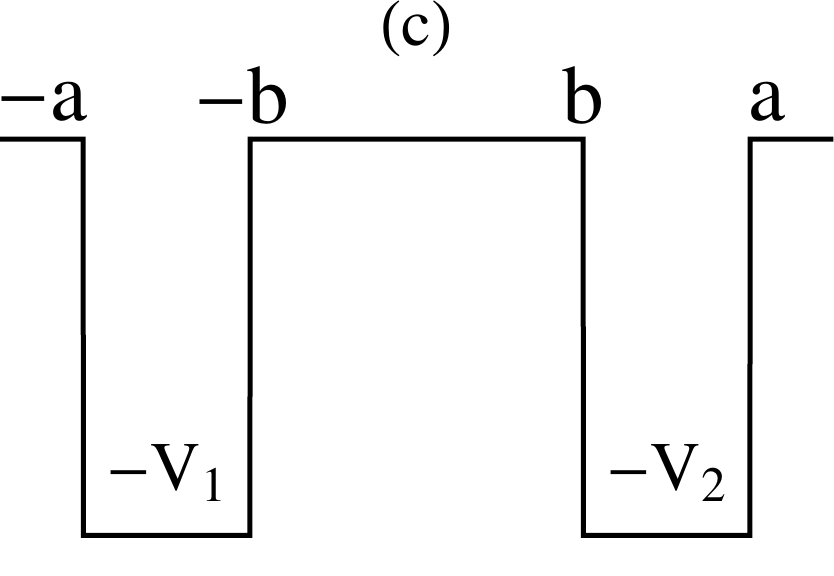}
\caption{Depiction of various double-well potentials perturbed by imaginary ($g \ne 0$) PT-symmetric potentials. Here, $V_1=u+ig, V_2=u-ig$. (a): double delta, (b): double delta between two rigid walls, (c):   square double-well. }
\end{figure}
In Fig. 2(a,b,c), we show the parametric evolution  of eigenvalues $E_n(g)$ for various PT-symmetric potentials [6,7] and for their Hermitian counterparts.
\begin{figure}[t]
	\centering
	\includegraphics[width=14 cm,height=6.3 cm]{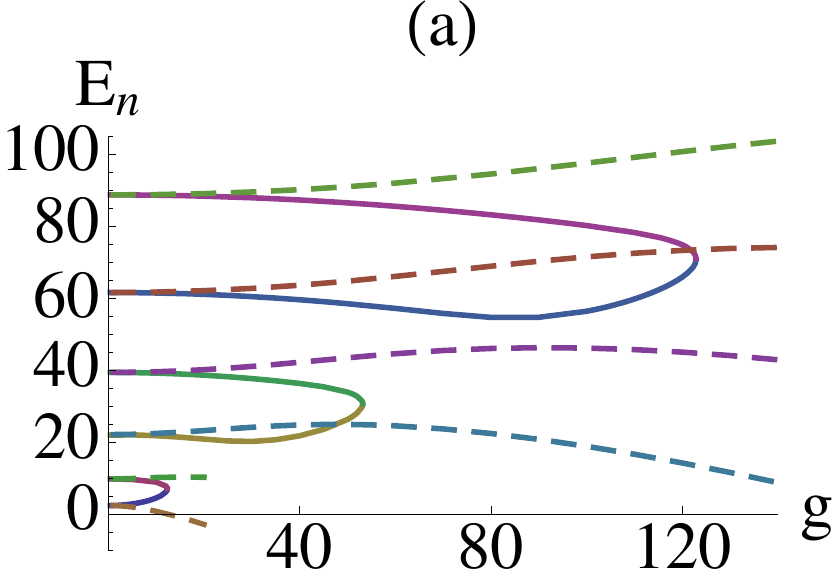}\\
	\includegraphics[width=14 cm,height=6.3 cm]{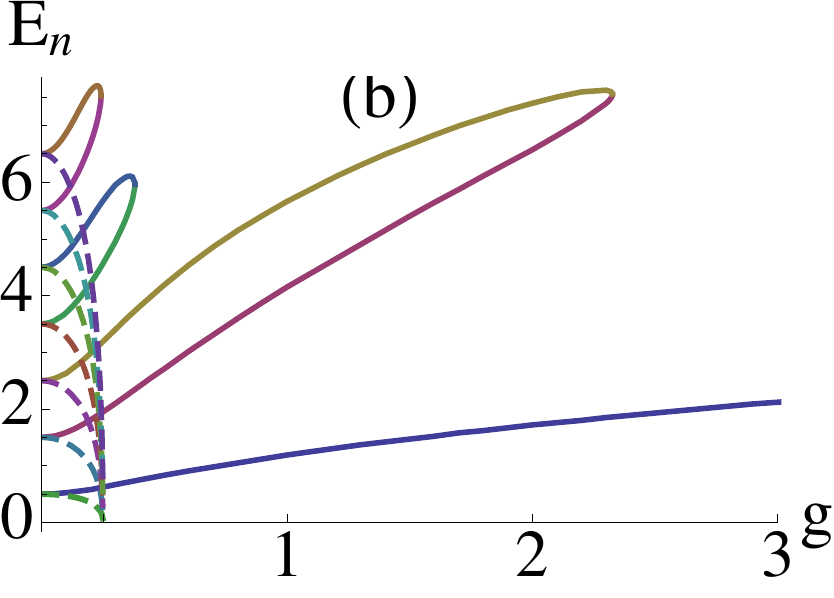}\\
	\includegraphics[width=14 cm,height=6.3 cm]{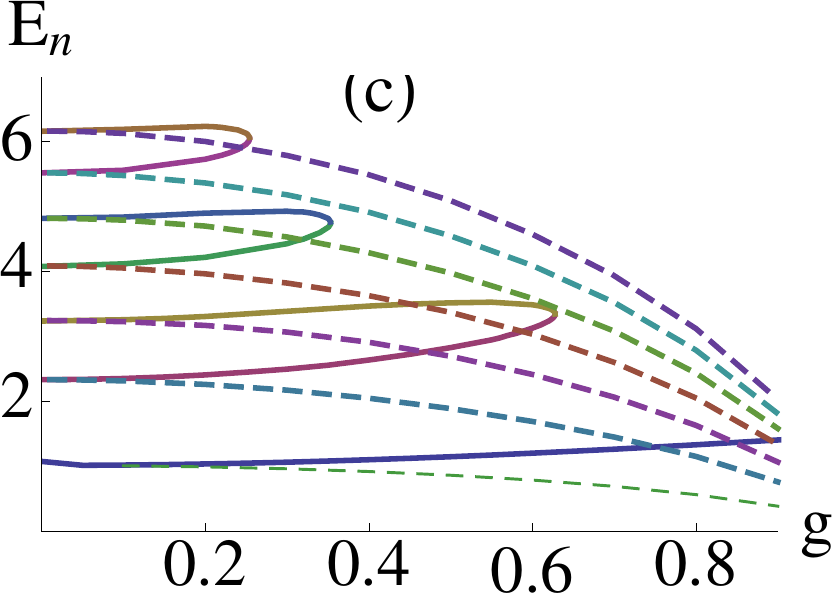}
	\caption{The scenario of coalescing of eigenvalues (solid curves) for complex PT-symmetric potentials. (a): $V(x)=igx$ [6] between two rigid walls at $x=\pm 1$, (b): $V(x)=x^2/4+igx|x|$ [7], (c): $V(x)=|x|+igx$ [7]. In all three case $E_n(-g)=E_n(g)$. The eigenvalues of their Hermitian counterparts (when $g$ is replaced by $\pm ig$) are shown by dashed lines. In (b) and (c) they exist for $|g|<1/4$ and $|g|<1$, respectively. Notice the qualitative dis-similarity between two types of spectra. However, quantitatively for small values of $g$ they do coincide.} 
\end{figure}
\begin{figure}[h]
\centering
\includegraphics[width=14 cm,height=6.5 cm]{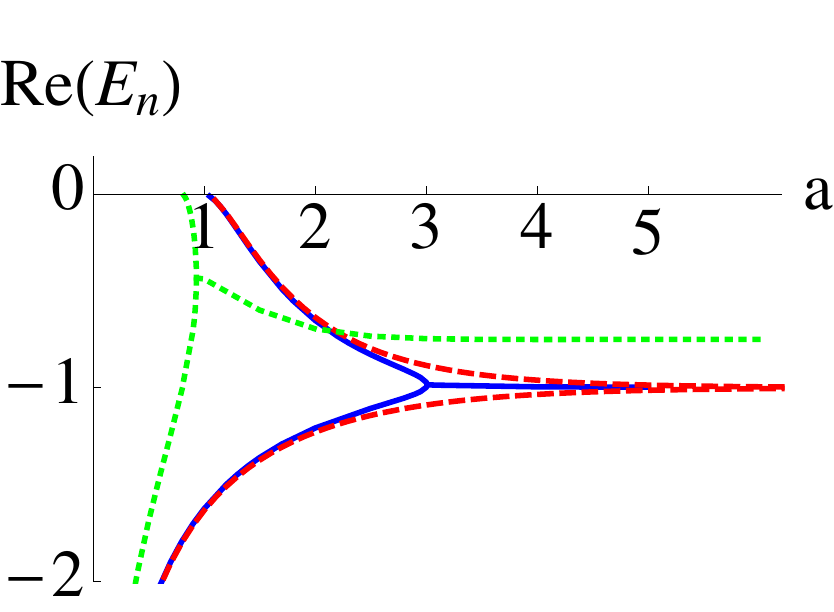}
\caption{Real part of $E_n$ of the two levels of the DDDP when $a$ is varied slowly: (i) $V_1=2=V_2$ (dashed line), (ii) $V_1=2+0.1i,V_2=2-0.1i$ (solid line), (iii) $V_1=2+i,V_2=2-i$ (dotted line). Dotted and solid lines represent coalescing of eigenvalues in complex PT-symmetric case. Dashed lines display merging of levels in Hermitian case. The eigenvalues are obtained from Eq. (1)}
\end{figure}
\begin{figure}[h]
\centering
\includegraphics[width=14 cm,height=6.5 cm]{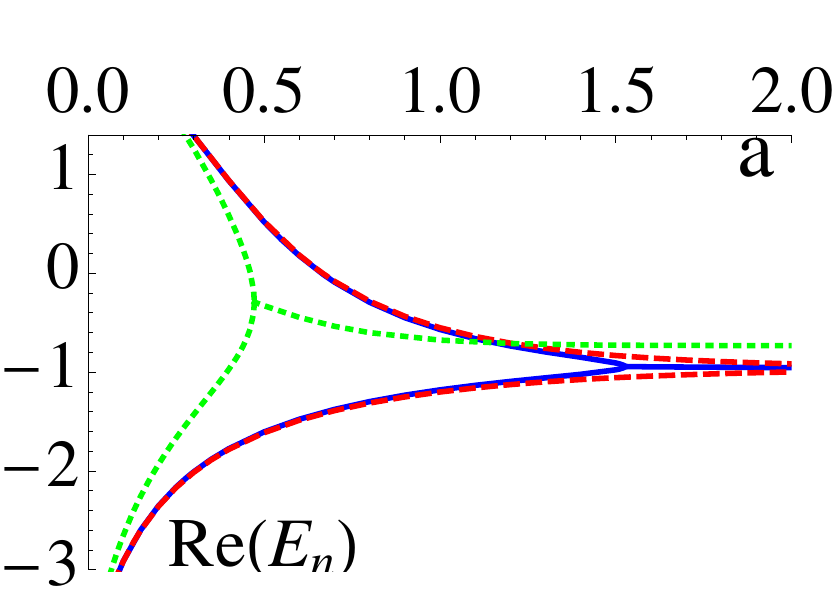}
\caption{Real part of $E_n$ of the two levels of the double delta well between two rigid walls (Fig 1(b)), when $a$ is varied slowly: (i) $V_1=2=V_2$ (dashed line), (ii) $V_1=2+0.1i,V_2=2-0.1i$ (solid line), (iii) $V_1=2+i,V_2=2-i$ (dotted line). Dotted and solid lines represent coalescing of eigenvalues in complex PT-symmetric case. Dashed lines display merging of levels in Hermitian case.  Not shown are the next two levels which are real without any coalescing. Eigenvalues have been obtained form Eq. (3)}
\end{figure}

\begin{figure}[h]
\centering
\includegraphics[width=14 cm,height=6. cm]{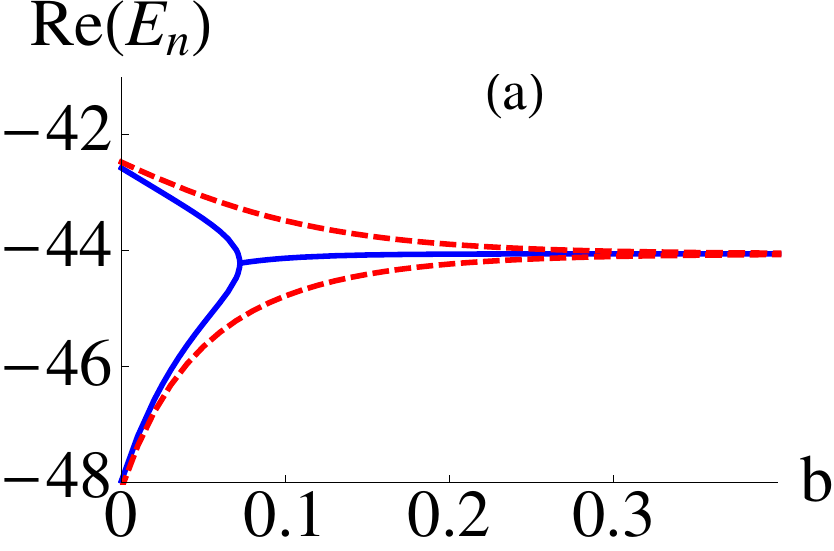}\\
\includegraphics[width=14 cm,height=6. cm]{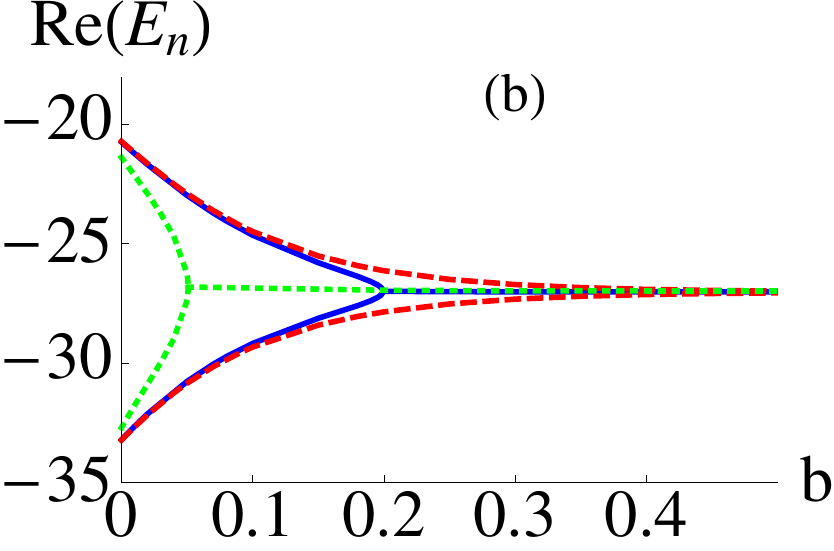}\\
\includegraphics[width=14 cm,height=6. cm]{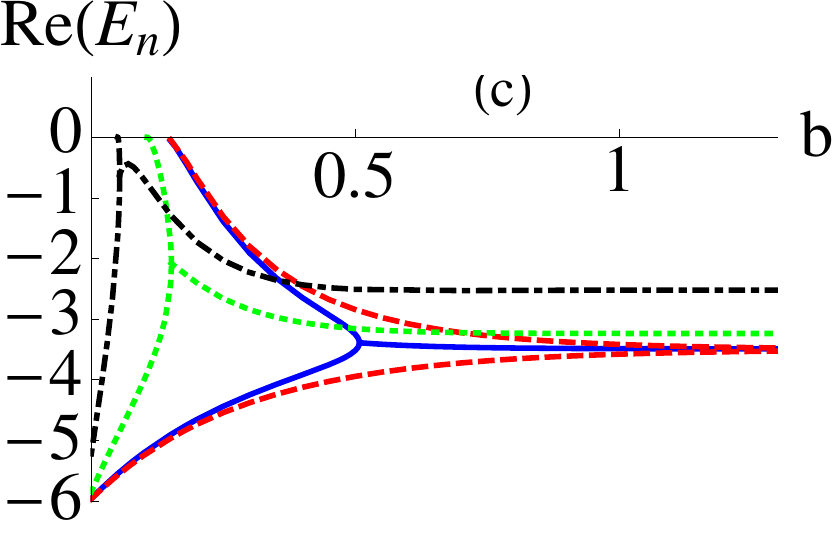}
\caption{Real part of $E_n$ of the six levels of the Square double-well (Fig. 1(c)) when $b$ is varied: (i) $V_1=50=V_2$ (dashed line), (ii) $V_1=50+i,V_2=50-i$ (solid line), (iii) $V_1=50+5i,V_2=50-5i$ (dotted line), (iv) $V_1=50+10i,V_2=50-10i$ (dot-dashed line).  Excepting red dashed lines, others represent coalescing of eigenvalues in complex PT-symmetric case . Red dashed lines present merging of levels in Hermitian case. (a): $E_0, E_1$ for $g=0,1$ (b): $E_2, E_3$ for $g=0,1$ and $E_0,E_1$ for $g=5$ (c): $E_4, E_5$ for $g=0,1$, $E_2,E_3$ for  $g=5$  and $E_0, E_1$ for  $g=10$. Eq. (8) has been used here.}
\end{figure}
For Double Dirac Delta Potential(DDDP, Fig. 1(a)) the formula for finding the eigenvalues of bound states [2] can be written as
\begin{equation}
4p^2-2pu=(u^2+g^2)(e^{-2pa}-1), \quad p=\sqrt{-2\mu E}/\hbar.
\end{equation} 

The solution of Schr{\"o}dinger equation for the Double Dirac-delta well between two rigid walls  (fig. 1(b)) is given by 
$\psi(-a<x<-b)= A \sinh p (x+a), \\ \psi(-b \le x \le b)=B e^{px} + C e^{-px}, \psi(b<x \le a)= D\sinh p(x-a)$.
\begin{eqnarray}
B e^{-pb}+Ce^{pb}=A\sinh pd, \quad (V_1+p)Be^{-pb}+(V_1-p)Ce^{pb}=Ap\cosh pd \\ \nonumber
Be^{pb}+Ce^{-pb}=-D\sinh pd, \quad (V_2-p)Be^{pb}+(V_2+p)Ce^{-pb}=-Dp\cosh pd
\end{eqnarray}
\begin{eqnarray}
e^{2pb}[V_1-p(1+\coth pd)][V_2-p(1+\coth pd)]=e^{-2pb}[V_1+p(1-\coth pd)][V_2+p(1-\coth pd)].
\end{eqnarray} 
Here, $V_1=u+ig, V_2=u-ig, d=a-b.$

The solution of Schr{\"o}dinger equation  for the square double-well potential (Fig.1(c))
is given as, $\psi(x<-a)= A e^{px} + B e^{-px}, \psi(-a \le x \le -b)= C e^{iqx} + D e^{-iqx}, \psi(-b<x<b)= F e^{px} + G e^{-px}, \psi(b \le x \le a)= H e^{irx} + K e^{-irx}$ and  $\psi(x>a)= L e^{px} + M e^{-px},$ where, 
\begin{equation}
 q=\sqrt{2\mu(E+u+ig)}/\hbar, \quad r=\sqrt{2\mu(E+u-ig)}/\hbar.
\end{equation}
\begin{eqnarray}
A e^{-pa}+Be^{pa}=C e^{-iqa}+D e^{iqa}, \quad A p e^{-pa}-B pe^{pa}=iCq e^{-iqa}-iD q e^{iqa} \\ \nonumber
C e^{-iqb}+D e^{iqb}=F e^{-pb}+G e^{pb}, \quad iCq e^{-iqb}-iDq e^{iqb}=Fp e^{-pb}-Gp e^{pb} \\ \nonumber
F e^{pb}+G e^{-pb}=H e^{irb}+K e^{-irb}, \quad Fp e^{pb}-Gp e^{-pb}=iHr e^{irb}-iKr e^{-irb} \\ \nonumber
H e^{ira}+K e^{-ira}=L e^{pa}+M e^{-pa}, \quad iHr e^{ira}-iKr e^{-ira}=Lp e^{pa}-Mp e^{-pa} 
\end{eqnarray}
\begin{eqnarray}
M_1 \left(\begin{array}{clcr} A \\B \end{array}\right)= M_2 \left(\begin{array}{clcr}C \\D \end{array}\right), \quad M_3 \left(\begin{array}{clcr} C\\D\end{array}\right)=M_4 \left(\begin{array}{clcr} F \\G \end{array}\right) \\ \nonumber M_5 \left(\begin{array}{clcr} F \\G \end{array}\right)=M_6 \left(\begin{array}{clcr} H \\K \end{array}\right), \quad M_7 \left(\begin{array}{clcr} H \\K \end{array}\right)=M_8 \left(\begin{array}{clcr} L \\M \end{array}\right)
\end{eqnarray}
\begin{equation}
\left(\begin{array}{clcr} A \\B \end{array}\right)= M^{-1}_1  M_2  M^{-1}_3  M_4  M^{-1}_5  M_6  M^{-1}_7  M_8 \left(\begin{array}{clcr} L \\M \end{array}\right)=\left(\begin{array}{clcr} m_{11}(E) & m_{12}(E) \\ m_{21}(E) & m_{22}(E) 
\end{array}\right)\left(\begin{array}{clcr} L \\M \end{array}\right)
\end{equation}
For bound states, we demand $B=0=L$. From Eq.(7) we have $B=L~m_{21}(E)+M~m_{22}(E)$, where $M\ne 0$. Finally we get, \begin{equation}
m_{22}(E)=0
\end{equation} 
gives the eigenvalues of bound states of double-well in Fig 1(c). 
Using Eqs. (1,3,6), we obtain eigenvalues of the double-well potentials (Fig. 1(a,b,c)) for both cases $g=0$ and $g\ne 0$.

In Fig 2, by solid curves, we show the evolution of eigenvalues $E_n(g)$ for three PT-symmetric potentials. By dashed curves, we 
show the same for Hermitian counterparts of these potentials when $g$ is replaced by $\pm ig$.
Notice the coalescing of eigenvalues at special values of $g$, these special values are called
EPs. Also note that at or around the EPs, the evolution of  eigenvalues in dashed lines does not show any special feature.  So the spectra of Hermitian and their complex PT-symmetric counterparts do not relate to each other well, excepting for very small values of $g$ where the eigenvalues of both coincide approximately.

In Fig. 3-5, we present the variation of eigenvalues when the distance between wells in Fig. 1(a,b,c) is increased slowly. The blue lines present the coalescing of two levels when the total potential is mildly complex PT-symmetric ($g$ is small). See the dashed red lines for Hermitian double-well ($g=0$) representing merging of two levels. Green dotted and blue dot-dashed  lines
arise when non-Hermiticity parameter becomes large, then the coalesced levels  are not contained between the merging levels (red dashed lines).

Lastly, we conclude that the models discussed here bring the spectral phenomena of coalescing and merging of energy levels closer, however we  know that they occur in two different domains: Hermitian and complex PT-symmetric. Further investigations in this regard are welcome.

\end{document}